\documentclass[12pt]{article}
\usepackage{amsmath}
\usepackage{epsfig}
\begin{document}
\title{ High energy photoionization of fullerenes}
\author{E. G. Drukarev, A. I. Mikhailov\\
{\em National Research Center "Kurchatov Institute"}\\
{\em B. P. Konstantinov Petersburg Nuclear Physics Institute}\\
{\em Gatchina, St. Petersburg 188300, Russia}}

\date{}
\maketitle

\begin{abstract}
{We show that the theoretical predictions on high energy behavior of the photoionization cross section of fullerenes depend crucially on the form of the function $V(r)$ which approximates the fullerene field. The shape of the high energy cross section is obtained without solving  the wave equation. The cross section energy  dependence is determined by the analytical properties of the function $V(r)$.}
\end{abstract}

\section{Introduction}
In this paper we calculate the high energy nonrelativistic asymptotics for the photoionization cross section of the valence electrons of fullerenes $C_N$. We consider the fullerenes which can be treated approximately as having the spherical shape. The photon carries the energy $\omega$ which is much larger than the ionization potential $I$. We find the leading term of the cross section expansion in terms of $1/\omega$.  We keep the photon energy to be much smaller than the electron rest energy $mc^2$. Here we consider only the ionization of $s$ states. We employ the relativistic system of units in which $ \hbar=1$; $c=1$
and the squared electron charge $e^2=\alpha=1/137$.

The actual potential experienced by the fullerene valence electrons is a
multicentered screened Coulomb potential produced by the $C^{+4}$ carbon ions of the
fullerene. The ionized electron approaches one of these centers transferring large momentum.
The asymptotic of the photoionization cross section $\sigma(\omega)$ in the screened Coulomb field
is the same as in the unscreened one \cite{DM}, i.e. $\sigma \sim \omega^{-7/2}$.
Thus we expect the observed asymptotic also to be $\sigma \sim \omega^{-7/2}$.

However one usually uses a model central potential $V(r)$ for description of the field created by the fullerene. For a model potential the asymptotics may be a different one. We consider spherical fullerene with the radius $R$ and the width of the layer $\Delta\ll R$. The general properties of the potential $V(r)$ are well known--see, e.g. \cite{1}. It is located mostly inside the fullerene layer $R -\Delta/2 \leq r \leq R+\Delta/2$ being negligibly small outside.

In the simplest (or even the oversimplified \cite{2}) version it is just the well potential
which is constant inside the layer and vanishes outside. Introducing $R_{2,1}=R \pm \Delta/2$ we can present the potential as
\begin{equation}
V(r)\ =\ -V_0\theta(r-R_1)\Big(1-\theta(r-R_2)\Big); \quad V_0\ >\ 0.
\label{1}
\end{equation}
Recall that $\theta(x)=1$ for $x \geq 0$ while $\theta(x)=0$ for $x<0$.
One often uses
the Dirac bubble potential
\begin{equation}
V(r)\ =\ -U_0\delta(r-R),
\label{2}
\end{equation}
in the fullerene studies \cite{3}.  Here, as well as in Eqs.(3) and (4) $U_0>0$ are the dimensionless constants.

The nowadays calculations are often based on the jellium model \cite{3a} (see, e.g.\cite{3b},\cite{3c}). In this approach the charge of the positive core consisting of nuclei and the internal electrons is assumed to be distributed uniformly in the fullerene layer. The field of the positive core is $V(r)=V_1(r)$ at $0\leq r <R_1$, $V(r)=V_2(r)$ at $R_1\leq r  \leq R_2$, and $V(r)=V_3(r)$ at $ r > R_2$ with
\begin{equation}
V_1(r)\ =\ const=-U_0\frac{3}{2}\frac{R_2^2-R_1^2}{R_2^3-R_1^3} ; \quad V_2(r)=-\frac {U_0}{2(R_2^3-R_1^3)}\Big(3R_2^2-r^2(1+\frac{2R_1^3}{r^3})\Big);
\label{2a}
\end{equation}
$$V_3(r)=-\frac{U_0}{r}.$$

Sometimes model potentials are determined by analytical functions of $r$ with a sharp peak at $r=R$. The Lorentz bubble potential is
\begin{equation}
V(r) = -\frac{U_0}{\pi}\frac{a}{(r-R)^2+a^2}.
\label{3}
\end{equation}
It describes the Dirac bubble potential at
$ a \rightarrow 0$. The Gaussian-type potential
\begin{equation}
V(r)\ =\ -\frac{V_0}{\pi}\exp{\frac{-(r-R)^2}{a^2}},
\label{4}
\end{equation}
with $a \approx \Delta \ll R$ was employed in \cite{4}.

Strictly speaking our analysis is true for the negative ion $C^{-}_N$. However, since there are many valence electrons in the fullerene shell, we expect it to be true for photoionization of the neutral fullerene $C_N$ as well.

As it stands now, the asymptotics for the photoionization cross section is known only for the Dirac bubble potential \cite{5}.
Here we demonstrate that the energy behavior of the asymptotic cross section is strongly model dependent. It is determined by the analytical properties of the potential $V(r)$. In Sec. 2 we obtain the general equation for the asymptotics of the photoionization cross section.
In Sec.3 we calculate the asymptotics for the potentials mentioned in Introduction. We analyze the results in Sec.4.

\section{Asymptotics of the cross section}

The photoionization cross section can be presented as (see Eq.(56.3) of \cite{4a} or Eq.(5.76) of \cite{3})
\begin{equation}
d\sigma=n_e\frac{mp}{(2\pi)^2}|F|^2d\Omega.
\label{5}
\end{equation}
Here $m$ is the electron mass, $p=|{\bf p}|$, while ${\bf p}$ is the photoelectron momentum, $\Omega$ is the solid angle of the photoelectron, and $n_e$ is the number of electrons in the ionized state. The normalization factor of the photon wave function
$n(\omega)=\sqrt{4\pi}/\sqrt{2\omega}$ is included in the photoionization amplitude $F$. Averaging over polarizations of the incoming photon is assumed to be carried out.

We consider the photon energy $\omega$ which is much larger than the ionization potential $I$, i.e. $\omega \gg I$. Limiting  ourselves by the condition $\omega \ll m$ we can treat the photoelectron in nonrelativistic approximation. The kinetic energy of the photoelectron is $\varepsilon=\omega-I=p^2/2m$. The electron momentum $p$ is much larger than  the characteristic momentum $\mu=(2mI)^{1/2}$ of the bound state ($p \gg \mu$). At $\omega \gg I$ the photoionization requires large momentum ${\bf q}={\bf k}-{\bf p}$ to be transferred to the recoil fullerene. Here ${\bf k}$ is the photon momentum, and $k=|{\bf k}|=\omega$.
One can see that $k \ll p$ if $I \ll \omega \ll m$ and thus we can put $|{\bf q}|=q=p$.

If the electron-photon interaction is written in the velocity form, momentum $q$ is transferred in the initial state in ionization of $s$ states \cite{6}, \cite{3}. Thus
the photoelectron can be described by plane wave. Interaction of the photoelectron with the ionized fullerene provides the contributions of the relative order $O(1/p)$ to the amplitude. Hence they contribute to the cross section beyond the asymptotics.  The photoionization amplitude can be written as
$ F=n(\omega)\int d^3r \psi^*_{\bf p}({\bf r})\gamma\psi(r)$ with $\psi_{\bf p}$ and $\psi$ the wave functions of the photoelectron and the bound electron correspondingly; $\gamma=-i\sqrt{\alpha}{\bf e}\cdot{\bf\nabla}/m$ is the operator of interaction between the photon and electron. In momentum space the amplitude takes the form
$$F=\sqrt{\alpha}n(\omega)\int \frac{d^3f}{(2\pi)^3}\psi_{\bf p}({\bf f})\frac{{\bf e}\cdot{\bf f}}{m}\psi({\bf f}-{\bf k}).$$
Since the photoelectron is described by the plane wave, i.e.  $\psi_{\bf p}({\bf f})=(2\pi)^3\delta({\bf f}-{\bf p})$,
the amplitude of photoionization can be written as \cite{6}
\begin{equation}
F=N(\omega)\frac{{\bf e}\cdot {\bf p}}{m}\psi(p); \quad N(\omega)=\Big(\frac{4\pi\alpha}{2\omega}\Big)^{1/2}.
\label{5a}
\end{equation}
We replaced $q$ by $p$ in the argument of the Fourier transform of the wave function of the fullerene electron.
The latter is $\psi(p)=\int d^3r \psi(r)e^{-i{\bf p}\cdot{\bf r}}$.

Now we present the wave function $\psi(p)$
in terms of the Fourier transform of the potential
\begin{equation}
V(p)=\int d^3r V(r)e^{-i{\bf p}\cdot{\bf r}}=\frac{4\pi}{p}\int _0^{\infty} dr rV(r)\sin{pr}.
\label{6}
\end{equation}

The function $\psi(p)$ can be expressed by the Lippmann--Schwinger equation \cite{3}
\begin{equation}
\psi=\psi_0+G(\varepsilon_B)V\psi,
\label{7}
\end{equation}
with $G$ the electron propagator of free motion, $\varepsilon_B=-I$ is the energy of the bound state. The matrix element of the propagator is
$$ \langle {\bf f}_1|G(\varepsilon_B)|{\bf f}_2\rangle=g(\varepsilon_B, f_1)\delta({\bf f}_1-{\bf f}_2); \quad g(\varepsilon_B, f_1)=\frac{1}{\varepsilon_B-f_1^2/2m}.$$
For a bound state $\psi_0=0$, and thus Eq.(\ref{7}) can be evaluated as
$\psi(p)=\langle {\bf p}|GV|\psi\rangle=g(\varepsilon_B, p)J(p)$ with
\begin{equation}
J(p)=\int\frac{d^3f}{(2\pi)^3}\langle {\bf p}|V|{\bf f}\rangle\langle{\bf f}|\psi\rangle=\int\frac{d^3f}{(2\pi)^3}V({\bf p}-{\bf f})\psi({\bf f}).
\label{7a}
\end{equation}
Putting also $g(\varepsilon_B, p)=-2m/p^2$ we obtain
\begin{equation}
\psi(p)=-\frac{2m}{p^2}J(p)=-\frac{J(p)}{\omega}.
\label{8}
\end{equation}
The integral $J(p)$ is saturated at $f \sim \mu \ll p$. Thus its dependence on $p$ is determined by that of $V(p)$.
Another presentation
\begin{equation}
J(p)=\int d^3r \psi(r)V(r)e^{-i{\bf p}\cdot{\bf r}}=\frac{4\pi}{p}\int_0^{\infty}dr\chi(r)V(r)\sin{(pr)}; \quad \chi(r)=r\psi(r),
\label{7b}
\end{equation}
can be obtained by the Fourier transformation of the integrand on the right hand side of Eq.(\ref{7a}).

We shall demonstrate that the potential can take the form
$V(p)=V_1(p)+V_2(p)$ with the two terms corresponding to two fullerene characteristics $R_1$ and $R_2$.
In this case
$J(p)$ can be presented as
\begin{equation}
J(p)=V_1(p)\kappa_1+V_2(p)\kappa_2,
\label{3xy}
\end{equation}
where the factors $\kappa_{1,2}$ do not depend on $p$, being determined by the characteristics of the bound state.
Thus Eq. (\ref{8}) can be written as
\begin{equation}
\psi(p)=-\frac{1}{\omega}\Big(V_1(p)\kappa_1+V_2(p)\kappa_2\Big),
\label{8N}
\end{equation}
and Eq.(\ref{5a}) can be presented  as
\begin{equation}
F=-\frac{N(\omega)}{\omega}\frac{{\bf e}\cdot {\bf p}}{m}\sum_{i=1,2}V_i(p)\kappa_i.
\label{9}
\end{equation}
Hence the asymptotics of the photoionization cross section can be expressed through the potential $ V(p)$:
\begin{equation}
\sigma(\omega)=\frac{4\alpha}{3}\frac{p}{\omega^2}n_e|\sum_iV_i(p)\kappa_i|^2; \quad p=\sqrt{2m\omega}.
\label{10}
\end{equation}

If the potential is determined by analytical function of $R$ and $\Delta$, we have just $J(p)=V(p)\kappa$. It was shown in \cite{DM} that in the simplest case $\kappa=\psi(r=0)$.

Thus one can find the asymptotic energy dependence of the photoionization cross section without solving the wave equation.

\section{Asymptotics for the model potentials}

Start with the well potential defined by Eq.(1). Employing Eq.(\ref{7b}) we find immediately
\begin{equation}
\psi(p)=-\frac{4\pi V_0R}{\omega p^2}\lambda; \quad \lambda=\cos{(pR_2)}\psi(R_2)-\cos{(pR_1)}\psi(R_1).
\label{39}
\end{equation}
Here we neglected the terms of the order $\Delta/R \ll 1$.
The parameter $\lambda$ can be expressed in terms of the fullerene characteristics $R$ and $\Delta$.
%$$ \lambda=\cos{(pR)}\cos{(p\Delta/2)}\psi(R_2)-\sin{(pR)}\sin{(p\Delta/2)}\psi(R_1).$$
Note that the function $\psi(r)$ varies noticeably inside the fullerene layer $\Delta$. Hence the difference $\psi(R_2)-\psi(R_1)$ is not small.
The photoionization cross section is thus
\begin{equation}
\sigma=\frac{2^6\alpha\pi^2}{3}\frac{V_0^2R^2}{\omega^2 p^3}n_e\lambda^2.
\label{39a}
\end{equation}
It drops as $\omega^{-7/2}$.

For the Dirac bubble potential given by Eq.(2) we obtain employing Eq.(\ref{7b})
\begin{equation}
\psi(p)=\frac{4\pi U_0R}{\omega p}\sin{(pR)}\psi(R),
\label{50}
\end{equation}
Thus the cross section is
\begin{equation}
\sigma=\frac{2^6\alpha\pi^2}{3}\frac{U_0^2R^2}{\omega^2p}\sin^2{(pR)}n_e\psi^2(R); \quad p^2=2m\omega,
\label{50}
\end{equation}
dropping as $\omega^{-5/2}$. This is just the result obtained in \cite{5} by solving the wave equation.

Both Dirac bubble and well potentials have singularities on the real axis. However, the wave functions exhibit different behavior at the singular points. This leads to different high energy behavior in these cases. The wave function corresponding to the Dirac bubble potential is continuous at $r=R$ while
 its first derivative suffers a jump at this point. This can be demonstrated by integration the wave equation
 \begin{equation}
 \chi^{(2)}(r)=2mV(r)\chi(r)-2m\varepsilon_B\chi(r),
 \label{50n}
\end{equation}
with $V(r)$ determined by Eq.(\ref{2}) over the small interval $R^{-} \leq r \leq R^{+}$ with $R^{\pm}=R\pm\delta, \delta \rightarrow 0$.

The wave function of the bound electron in momentum representation can be expressed in terms of these jumps. After two integrations by parts of the expression
$$\psi(p)=\frac{4\pi}{p}\int_0^{\infty}dr\sin{(pr)}\chi(r),$$
we find for the Dirac bubble potential
\begin{equation}
\psi(p)=\frac{4\pi}{p^3}R\sin{(pR)}[\psi'(R^{-})-\psi'(R^{+})]
\label{51}
\end{equation}
$$-\frac{4\pi}{p^3}\Big[\int_0^{R^{-}}dr\sin{pr}\chi^{(2)}(r)+\int_{R^{+}}^{\infty}dr\sin{pr}\chi^{(2)}(r)\Big]$$
Further integration by parts of the second term demonstrates that it is about $1/p$ times the first one. Hence the leading contribution is provided by the first term and the Fourier transform of the function $\psi(r)$ is determined by the jump of
the first derivative. This leads to the $\omega^{-5/2}$ law for the cross section.

In the case of the well potential one can present the function $\psi(p)$ in similar form, but with two singular points $r=R_{2,1}=R\pm\Delta/2$. The first derivatives $\psi'(r)$ are continuous at these points while the second derivatives experience jumps \cite{Sh}. After three integrations by parts we find
$$\psi(p)=-\frac{4\pi}{p^4}\sum_{i=1,2}R_i\cos{(pR_i)}\delta\psi^{(2)}(R_i)$$
where $\delta\psi^{(2)}(R_i)=\psi^{(2)}(R_i+\delta)-\psi^{(2)}(R_i-\delta)$ are the jumps of the second derivative $\psi^{(2)}(R_i)$.
Thus the asymptotic wave function in the field of the well potential is $1/p$ times that in the Dirac bubble potential. This provides additional small factor of the order $1/\omega$ in the photoionization cross section given by Eq.(\ref{39a})compared with that for the Dirac bubble potential presented by Eq.(20).

The potential $V(r)$ for the jellium model given by Eq.(\ref{2a}) is continuous at the real axis as well as its derivative $V'(r)$.  One can see that $V_1(R_1)=V_2(R_1)$ and
$V_2(R_2)=V_3(R_2)$. Also $V_1'(R_1)=V_2'(R_1)$, $V_2'(R_2)=V_3'(R_2)$. The second derivative $V^{(2)}(r)$ experiences jumps at $r=R_{1,2}$.
Employing Eq.(\ref{7b}) we find after three integrations by parts
\begin{equation}
\psi(p)=\frac{4\pi}{\omega p^4}\lambda; \quad \lambda=\sum_{i=1,2}R_i\cos{(pR_i)}\psi(R_i)\delta V^{(2)}(R_i); \quad \delta V^{(2)}(R_i)=V^{(2)}(R_i^{+})-V^{(2)}(R_i^{-}).
\end{equation}
Note that the second derivatives $\psi^{(2)}(r)$ are continuous at $r=R_{1,2}$. This can be seen from the wave equation (21).
The cross section is
\begin{equation}
\sigma=\frac{64\alpha \pi^2}{3\omega^2p^7}n_e\lambda^2.
\label{58}
\end{equation}
It drops as $\omega^{-11/2}$.

Turn now to the Lorentz bubble potential determined by Eq.(\ref{3}) with  $a \ll R$. Employing Eq.(\ref{6}) and changing the variable of integration $x=r-R$ we present the  Fourier transform of the potential as
$$V(p)=V_A(p)+V_B(p); \quad V_{A,B}=-4\frac{a}{p}X_{A,B},$$
with
\begin{equation}
X_A(p)=U_0(-\frac{\partial}{\partial p})\cos{(pR)}\int_{-\infty}^{\infty}dx\cos{(px)}u(x);
\label{18}
\end{equation}
$$X_B(p)=U_0\frac{\partial}{\partial p}\int_{R}^{\infty}dx\cos{((p(x-R))}u(x) \quad u(x)=\frac{1}{x^2+a^2}.$$
Note that $X_A(p)$ is determined by $r$ close to $R$, i.e. by $|R-r|\sim a$.
Introducing $r'=x-R$ one can write the second equality as
$$X_B(p)=-U_0\int_{0}^{\infty}dr'r'\sin{(pr')}u(R+r').$$
It is dominated by small $r' \sim 1/p$.

Since the potential $V_A$ is determined by the space region where the electron density reaches it largest values, we expect it to be the most important one. Neglecting for a while the contribution of $V_B$ we find in the lowest order of expansion in powers of $a/R$
\begin{equation}
X_A(p)=\pi \frac{U_0R}{a}e^{-pa}\sin{pR}; \quad V(p)=-4\pi \frac{U_0R}{p}e^{-pa}\sin{pR}.
\label{64}
\end{equation}

We calculate the function $\psi(p)$ by employing Eqs.(\ref{7a}) and (\ref{8}). The potential $V(v)=-4\pi U_0Re^{-va}\sin{(vR)}/v$, with $v=|{\bf p}-{\bf f}|$ can not be expanded in powers of $f$.  Note, however  that it is sufficient to put $v=p-{\bf p}\cdot{\bf f}/p$ for calculation of the asymptotics of the wave function. Thus we obtain
\begin{equation}
\psi(p)=4\frac{\pi U_0R}{\omega}\int\frac{d^3f}{(2\pi)^3}\frac{e^{-va}}{v}[\sin {(pR)}\cos{(ftR)}-\cos{(pR)}\sin{(ftR)}]\psi(f); \quad t={\bf p}\cdot{\bf f}/pf.
\label{67}
\end{equation}
Due to the factor in the square brackets the integral over $f$ on the right hand side of Eq.(\ref{67}) is saturated by small $|ft| \sim 1/R \ll 1/a$.
This enables to put $v=p$ and $e^{-(p-ft)a}=e^{-pa}$ in the integrand.
Since the wave function $\psi(f)$ does not depend on the angular variables, we find immediately
\begin{equation}
\psi(p)=4\frac{\pi U_0R}{\omega p}e^{-pa}\sin {(pR)}A; \quad A=\frac{4\pi}{R}\int_0^{\infty}\frac{dff}{(2\pi)^3}\sin{(fR)}\psi(f).
\label{68}
\end{equation}
Noting that
$A=\psi(R)$, we obtain
\begin{equation}
\psi(p)=4\frac{\pi U_0R}{\omega p}e^{-pa}\sin {(pR)}\psi(R).
\label{69}
\end{equation}
The cross section
\begin{equation}
\sigma_A=\frac{64}{3}\frac{\alpha\pi^2U_0^2R^2}{\omega^2p}e^{-2pa}\sin^2{(pR)}n_e\psi^2(R),
\label{72}
\end{equation}
differs from that for the Dirac bubble potential given by Eq.(\ref{50}) by the exponential factor $e^{-2pa}$. Note that the latter is caused by the poles $r=R \pm ia$
of the potential (\ref{3}) in the complex plane.

The lowest term of the asymptotic expansion of the function $X_B(p)$ \cite{7} provides the potential
$$ V_B(p)=\frac{16U_0a}{p(pR)^3}.$$
From the formal point of view, the corresponding cross section
\begin{equation}
\sigma_B=\frac{2^{10}\alpha}{3}\frac{U_0^2a^2}{R^6}\frac{\psi^2(r=0)}{\omega^2p^7},
\label{71}
\end{equation}
dropping as $\omega^{-11/2}$ is the true asymptotics of the photoionization cross section. However, e.g., at characteristic values $R=6 $a.u, $a=\Delta=1$ a.u. the cross section determined by Eq.(\ref{71}) becomes comparable with that determined by Eq.(\ref{72}) only at the photoelectron energies $\varepsilon \geq $ 5 keV. At these energies both cross sections $\sigma_A$ and $\sigma_B$ are more than $10^8$ times smaller than the cross section $\sigma_A$ at $\varepsilon =100$ eV. They have no chances to be observed. Thus only the cross section $\sigma_A$ is of physical interest.

Similar analysis can be carried out for the Gaussian-type potential given by Eq.(\ref{4}). Now $V(p)=V_A(p)+V_B(p)$ with $V_{A,B}(p)=-4V_0X_{A,B}(p)/p$.
The functions $X_A(p)$ and $X_B(p)$ are given by Eq.(\ref{18}) with $u(x)=e^{-x^2/a^2}$ and $U_0=1$. The observable cross section
\begin{equation}
\sigma=\frac{64\alpha \pi}{3}\frac{V_0^2R^2a^2}{\omega^2p}e^{-p^2a^2/2}\sin^2{(pR)}n_e\psi^2(R),
\label{78}
\end{equation}
is determined by the potential $V_A$.
The Gaussian drop changes to the power behavior $\sigma (\omega) \sim \omega^{-11/2}$ at the photoelectron energies of about $3$ keV.
Here the cross section is too small to be detected.

\section{Summary}

We found that the high energy behavior of the photoionization cross section of fullerenes depends on the form of the function $V(r)$
which is chosen for approximation of the fullerene field. We expressed the asymptotic cross section in terms of the Fourier transform $V(p)$ of the potential $V(r)$ without solving the wave equation.

The shape of the function $V(p)$
at large $p$ is known to be determined by the analytical properties of the function $V(r)$ \cite{8}.
Thus the latter determine the shape of the asymptotic cross section as well.

The three potentials presented by Eqs.(\ref{1})-(\ref{2a})have singularities on the real axis. In each case the cross section exhibits a power drop. The cross section decreases as $\omega^{-5/2}$ in the  Dirac bubble potential which turnes to infinity at $r=R$. It behaves as $\omega^{-7/2}$ in the well potential with the finite jumps of $V(r)$. The potential of jellium model is more smooth, and only the second derivatives $V^{(2)}(r)$ experience jumps. In this model the cross section drops as $\omega^{-11/2}$.

The Lorentz bubble potential given by Eq.(\ref{3}) has poles in the complex plane. In this case the observable cross section exhibits exponential drop $e^{-2pa}$.  The Gaussian type potential determined by Eq.(\ref{4})  with the essential singularity in the complex plane provides Gaussian drop of the photoionization cross section. In both cases these fast drops change to a slower power drop $\omega^{-11/2}$. However this takes place at the energies of several keV where the cross sections become unobservably small.

\end{document}